\documentclass[namedreferences]{kluwer}
\usepackage{epsfig}

\begin{document}

\begin{article}

\begin{opening}

\title{A New Photometric Look at M51}

\author{V. \surname{Vansevi\v{c}ius} \email{wladas@itpa.lt}}
\institute{Institute of Physics, Go\v{s}tauto 12, Vilnius 2600, Lithuania \\
National Astronomical Observatory, Osawa 2-21-1, Mitaka, Tokyo 181, Japan}

\begin{abstract}
A new technique is used to derive the extinction and age of populations in the interacting
galaxies NGC5194 \& NGC5195 from high $S/N$ multicolor photometric data. A new evolutionary
scenario of the interaction is proposed.
\end{abstract}


\end{opening}

\section {Introduction}

Detailed investigation of populations in Local Group (LG) galaxies using stellar photometry
has recently become very popular (Aparicio 1998). This
is a direct way to study the populations and evolution of galaxies. However, even
for LG objects it is difficult to reach the required photometry depth.
To reconstruct the star formation history of more distant galaxies a different techniques
should be applied. Historically, the most popular method is an evolutionary population
synthesis (Tinsley 1980; Arimoto \& Yoshii 1986), which treats the entire galaxy as a
single zone unit. Such an approach is probably satisfactorily applicable only for ellipticals.
However, it seems too simplified in the case of spiral disks. To improve the tools available for the
evolutionary analysis of galaxies, a new method for multicolor mapping is introduced and
applied for the interacting galaxy pair NGC5194/95 (hereafter, M51 system).

M51 is a well studied nearby system of two overlapping galaxies seen nearly face-on, and
therefore suitable for testing new methods and $2-D$ mapping techniques. Since the
interpretation of the results of optical photometry always suffer from the influence
of dust residing in galaxies, it is important to know the extinction properties of the
galaxies under consideration. For this purpose, overlapping galaxies are particularly
suitable, allowing precise derivation of an attenuation law and the amount of the extinction in
the foreground galaxy (White \& Keel 1992). Moreover, results of an extensive dynamical
modeling of M51 are available (Barnes 1998), and hence can be used for a comparison
of the photometrically derived parameters with the ones determined by other independent
methods.

\section {Observations and Reductions}

Observations were carried out at Kiso Observatory (Japan) using the $1.05~m$ Schmidt
telescope on February 17 \& 18 and March 7, 1994. The CCD camera Tek 1024
(read-out noise $14~e^{-}$, scale $0.75~''/pixel$, and field of view $12.5' \times 12.5'$)
was used. The open cluster M67 was observed every night as a standard field. More than
150 exposures of M51 of $30-1800$ seconds duration and image quality of $3''-5''$ were
taken in total. The standard $BVR_{C}I_{C}$ and narrow-band
$H\alpha$ ($\lambda_{0}=658~nm; \delta\lambda=2.5~nm$) filters, together with
$Z$ ($\lambda_{0}=518~nm; \delta\lambda=21~nm$) filter of Vilnius system.

Standard reductions were made with IRAF. The $2-D$ aperture photometry (scanning the entire
field with a constant step of $\sim4''$ and aperture of $\sim8''$) was performed using
an AFO package (Brid\v{z}ius \& Vansevi\v{c}ius 1998).

Assuming the distance to the M51 system to be equal to $8.4~Mpc$
(Feldmeier, Ciardullo, \& Jacoby 1997), each circular $\sim8''$ aperture corresponds to the
region of $\sim330~pc$ in galaxies (hereafter, zone). In the color-magnitude diagram (CMD)
(Fig. 1) zones of both galaxies are shown separately. For further discussion we use
only precisely measured ($\sigma < 0.^{m}03$) zones with $V<23.^{m}5$.

It is worth noting the different morphologies and fine structures of both CMDs (Fig. 1).
Most of the "streaming features" are parallel to the reddening arrow, and can be attributed
to the dust influence.

\begin{figure}
%\centerline{\epsfig{file=figure1.eps,width=13cm}}
\caption{CMDs of the galaxies NGC5194 \& NGC5195. Galaxies are simply separated by declination.
Reddening arrow is given according to the M51 attenuation law.}
\end{figure}

\section{Extinction in M51}

The problem of dustiness of galaxies at different evolutionary stages
is widely discussed recently. In particular, the dilemma of whether the spiral disks are
optically thick or thin is debated (Davies \& Burstein 1995).
To clarify the problem of spiral disk dustiness, we determine the amount of
extinction and the attenuation curve in NGC5194 using NGC5195 as a
tracer. The facts supporting the validity of such a method are: a) NGC5195 is behind
NGC5194; b) NGC5195 is a source bright enough to have a high $S/N$ ratio even
in the most obscured parts of the galaxy; c) NGC5195 is found to be a regular S0 galaxy
in the near-infrared (Thronson, Rubin, \& Ksir 1991), therefore it
is possible to compare symmetrically located zones.

We measured two regions (symmetric with respect to the NGC5195 center and bar),
each consisting of 50 zones ($\sim20'' \times 40''$) at an average distance of $\sim40''$
from the center. Visual inspection of the $B$ image helped to locate the region
relatively free of dust signatures, and then the corresponding dusty region was
chosen on the opposite side of the center and bar. The $V$ and $B-V$ histograms of the
investigated zones are shown in Figure 2. The parameters of the derived attenuation curve are
given in Table I (definition of the extinction: $A_{V}=R_{V} \cdot E_{B-V}$, is used).
The derived median value of the extinction for the interarm region at $\sim11~kpc$ distance
from the center of NGC5194 is $A_{V}=0.7 \pm 0.3$.

\begin{figure}
%\centerline{\epsfig{file=figure2.eps,width=13cm}}
\caption{$V$ (panel (a): solid line - clear (without dust signatures) region, dashed line -
dusty region) \& $B-V$ (panel (b): no shading - clear region, shaded histogram - dusty region)
distributions of zones used to derive the attenuation law and extinction in the
disk of NGC5194.}
\end{figure}

\begin{table}[center]
\caption{Properties of the extinction in NGC5194.}\label{tab1}
\begin{tabular}{clccc}
\hline
 & Parameter & NGC5194 & Milky Way & \\
\hline
 & $R_{V}$            &  $2.56 \pm 0.13$  &  3.10  &  \\
 & $E_{V-R}/E_{B-V}$  &  $0.67 \pm 0.06$  &  0.56  &  \\
 & $E_{V-I}/E_{B-V}$  &  $1.33 \pm 0.08$  &  1.25  &  \\
 & $E_{B-I}/E_{B-V}$  &  $2.41 \pm 0.13$  &  2.25  &  \\
 & $E_{R-I}/E_{B-V}$  &  $0.66 \pm 0.09$  &  0.69  &  \\
 & $E_{Z-V}/E_{B-V}$  &  $0.30 \pm 0.05$  &  0.28  &  \\
\hline
\end{tabular}
\end{table}

Numerous determinations of the amount of extinction in NGC5194 have recently been obtained.
Smith et al. (1990) derived $A_{B}=0.9$ using NGC5195 as a tracing source.
Rix \& Rieke (1993) put a limit on the global optical depth of NGC5194 disk of
$\tau_{V}<1-2$. Nakai \& Kuno (1995) determined $A_{V}=1.16-4.38$, based on a study of
37 $HII$ regions in the disk of NGC5194.
Beckman et al. (1996) concluded that optical depth in the $V$ band, averaged out to a radius of
three scale lengths from the center of NGC5194, is 0.7. These results mainly
ascribe properties of the arm regions, however they do not contradict our conclusion
that the disk of NGC5194 is opaque in the interarm regions even at a large distance from the
center.

However, such a conclusion is in contradiction with the one drawn by
White, Keel, \& Conselice (1996) from a study of nine overlapping galaxy systems. They
derived the extinction $A_{B}=0.3-2.3$ in the arm and $A_{B}=0.08-0.47$ in the interarm
regions. NGC3314 is the only system in their sample which possesses significant extinction
in the arm ($A_{B}=1.11-1.64$) and in the interarm ($A_{B}=0.77-1.75$) regions, which
supports our conclusion that some of the spiral disks are opaque even in the interarm
regions.

The color excess ratios derived for the outer part of the disk are very close to those of
the Milky Way (MW) extinction law. Taking into account the same conclusion reported by
Panagia et al. (1996) for the central part of NGC5194, we assume that color excess ratios
are constant throughout the disk of this galaxy.

\section{Abundance and effective age of zones}

To avoid the age-abundance degeneracy in determining the effective age of zones, global
distribution of the abundance should be derived. Abundance analysis
of $HII$ regions (D\'{i}az et al. 1991; Zaritsky, Kennicutt, \& Huchra 1994) indicates
the mean abundance to be above solar: $12+log(O/H) \sim9.3$, with a gradient equal to
$\sim -0.06~dex/kpc$. Alternatively, Hill et al. (1997) explained the observed UV color
gradient in terms of radial extinction gradient, rather than of abundance.

We define an abundance parameter (independent of the extinction) as follows:
$Q_{ZVBV}=(Z-V)-(B-V) \cdot E_{Z-V}/E_{B-V}$.
$Q_{ZVBV}$ is a measure of $Mg~(518~nm)$ spectral feature strength, which is
a good abundance indicator (Barbuy 1994; Casuso et al. 1996). Plotting $Q_{ZVBV}$ vs.
radial distances from the galaxy centers, we do not see the abundance gradient except for the
very central region inside the radius of $\sim1.5~kpc$. However, a significant difference
of abundances between NGC5194 and NGC5195 is found. Therefore, deriving the ages of zones we
assume no abundance gradient across the disk of both galaxies. The abundances Z=0.05 for NGC5194
and Z=0.02 for NGC5195 are adopted, respectively. Such an assumption does not significantly affect
differential analysis of the age distribution of populations in the disk outside the inner
Lindblad resonance (radius $\sim 1.25~kpc$) of NGC5194 (Elmegreen, Elmegreen, \& Seiden 1989),
and outside the comparable radius in NGC5195.

Let us introduce a parameter $Q_{BVI}=(B-V)-(V-I) \cdot E_{B-V}/E_{V-I}$ as a measure of the
effective age of each zone, which is again extinction-independent. Actually, both
$Q$ formulae are only correct in the case of a star residing behind the dust cloud (screen). The
validity of dusty screen approximation in the case of face-on
galaxies was shown by Gonz\'{a}lez \& Graham (1996).
Diagram $V$ vs. $Q_{BVI}$ constructed from our data sample, in contrast to Figure 1, shows no
prominent "streaming features", which can be attributed to the influence of extinction.
All this indicates that $Q_{BVI}$ is a reddening-free parameter in the case of face-on galaxies.
Figure 3a shows calibration of $Q_{BVI}$ vs. age for single (age) stellar populations
(SSPs) calculated from the model spectra library GISSEL96 (Bruzual \& Charlot 1993). Figure 3b
provides histograms of the parameter $Q_{BVI}$ for both galaxies of the M51 system, which are
simply separated by declination. A significant difference in age between the two galaxies
is obvious even if the interfering influence of the Northern arm of NGC5194 projected onto
NGC5195 is neglected. The derived ages of the bulk of population in NGC5194 \& NGC5195 are
$390 \pm 70~Myr$ and $710 \pm 120~Myr$, respectively. The gray-coded population age
distribution in galaxies is shown in Figure 4.

\begin{figure}
%\centerline{\epsfig{file=fig3.eps,width=13cm}}
\caption{a) $Q_{BVI}$ vs. log(t) for solar (solid line) \& 2.5 times solar (dashed line)
abundances; b) $Q_{BVI}$ of NGC5194 (solid line) \& NGC5195 (dashed line).}
\end{figure}

\begin{figure}
%\centerline{\epsfig{file=fig4.eps,width=13cm}}
\caption{A map of the population ages ($Q_{BVI}$) in M51. Light gray corresponds to the
youngest and dark gray to the oldest zones.}
\end{figure}

Recent dynamic modeling suggests two starbursts in M51, the first less than $\sim100~Myr$ ago
and the second one $\sim400~Myr$ ago (see Barnes (1998) for discussion). We derive the age for
a bulk of zones in NGC5194 to be $\sim390~Myr$, in good agreement with the dynamical prediction.
Zones in NGC5195 are $\sim320~Myr$ older. If one assumes this age difference to be a
characteristic time between two close-up passages, the predicted age $\sim70~Myr$
for the last passage again is in good agreement with the dynamical modeling. Therefore,
our study indicates that the evolutionary scenario of M51 has had at least three periods of
activity. The first starburst occurred $\sim700~Myr$ ago in NGC5195
during a close-up passage with NGC5194, when, probably, NGC5195 was transformed to an S0 galaxy.
The next starburst took place after $\sim300~Myr$, when the bulk of the population in NGC5194 was
formed, and finally the last one occurred again after $\sim300~Myr$ when the present
grand-design appearance of NGC5194 was "touched up".

\section{Conclusions}

Main conclusions are the following: a) the attenuation law determined in NGC5194 is close to MW
standard extinction law; b) the opaque interarm region at distance $\sim11~kpc$ from the center
of NGC5194 supports optically thick models of spiral disks; c) the derived effective age
$\sim400~Myr$ for the bulk of NGC5194 zones agrees well with the predictions of recent dynamic
modeling of M51; d) the effective age $\sim700~Myr$ derived for the bulk of zones in NGC5195
implies a scenario with three periods of starburst activity. However, more sophisticated
radiative transfer models and realistic star-dust distributions as well as $2-D$ models of the
evolution of spiral disks, instead of simple SSP approach, are crucial for the interpretation
of the mapping data.

\begin{acknowledgements}
I am thankful to K. Kodaira and S. Yoshida for their help with an application for observing
time. Kiso observatory's staff is also generously acknowledged. I am indebted to
K. Zdanavi\v{c}ius for providing photoelectric measurements of M51 in Vilnius system.
Finally, I would like to thank A. Ku\v{c}inskas and Y. McLean for careful reading of
the manuscript.
\end{acknowledgements}

\end{article}

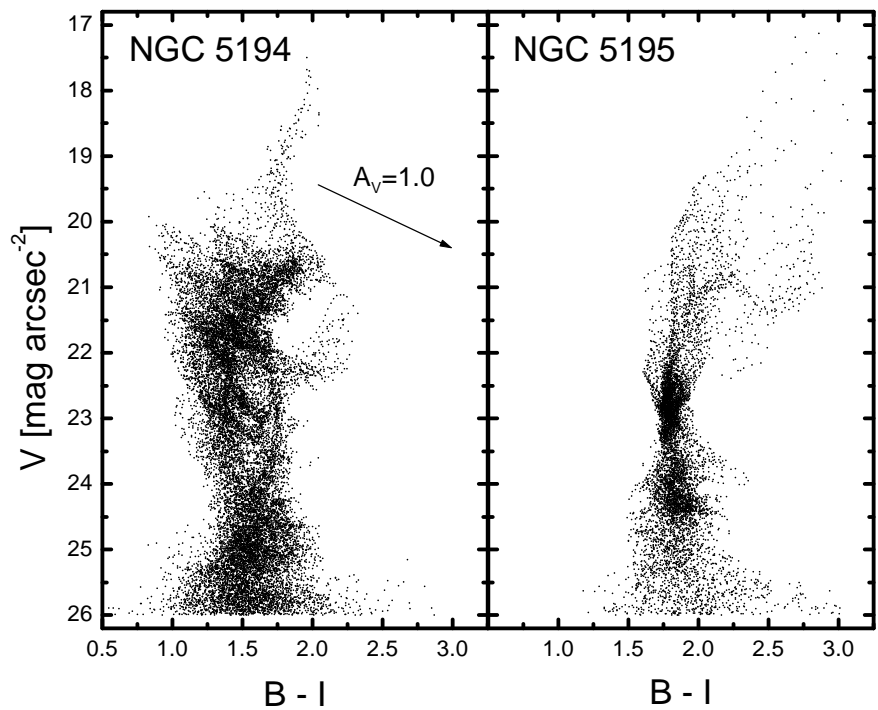
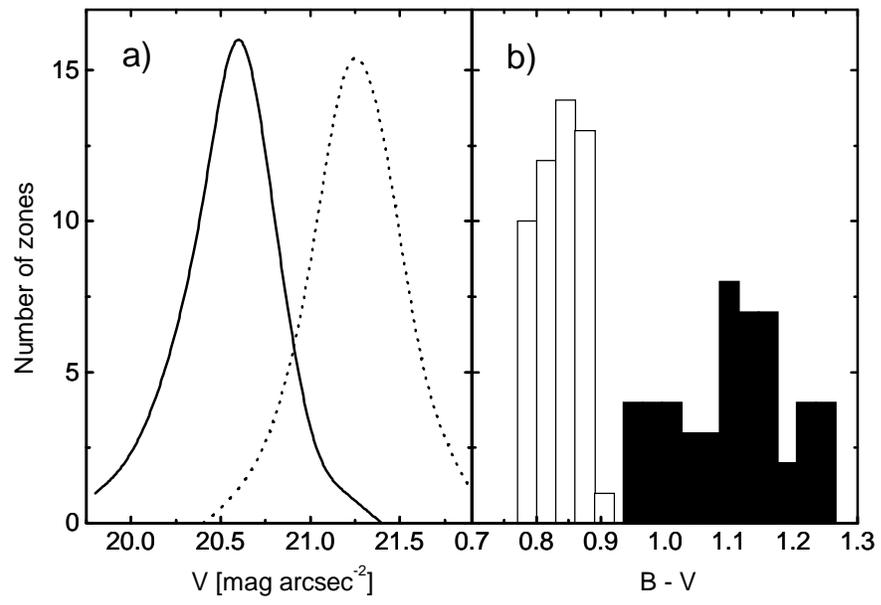
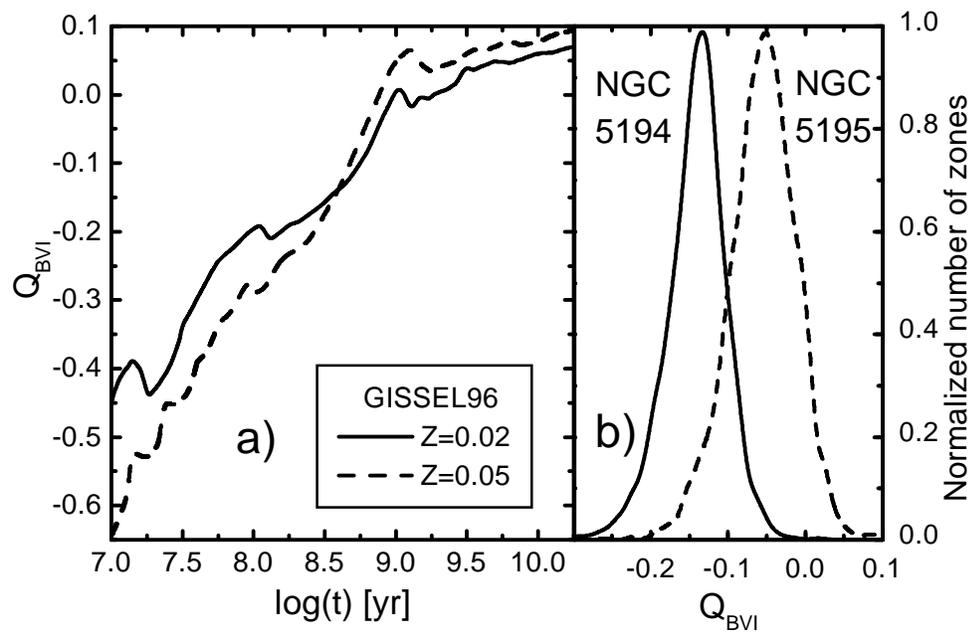
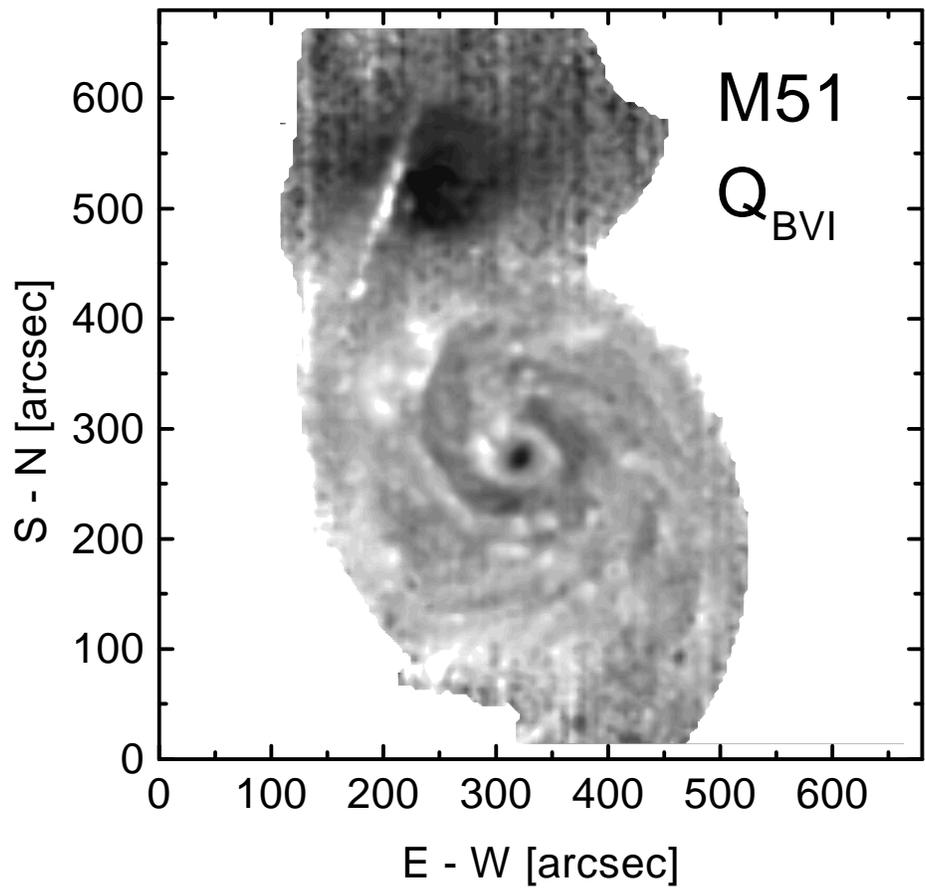
\end{document}